# The magnetoresistance and Hall effect in CeFeAsO: a high magnetic field study

H Q Yuan[1,*], L Jiao[1], J Singleton[2], F F Balakirev[2], G F Chen[3], J L Luo[3], N L Wang[3]

[1] Department of Physics, Zhejiang University, Hangzhou, Zhejiang 310027, China
[2] NHMFL, Los Alamos National Laboratory, MS E536, Los Alamos, NM 87545, USA
[3] Beijing National Laboratory for Condensed Matter Physics, Institute of Physics, Chinese Academy of Science, Beijing 10080, China

[*] E-mail: hqyuan@zju.edu.cn

**Abstract.** The longitudinal electrical resistivity and the transverse Hall resistivity of CeFeAsO are simultaneously measured up to a magnetic field of 45T using the facilities of pulsed magnetic field at Los Alamos. Distinct behaviour is observed in both the magnetoresistance $R_{xx}(\mu_0 H)$ and the Hall resistance $R_{xy}(\mu_0 H)$ while crossing the structural phase transition at $T_s \approx 150K$. At temperatures above $T_s$, little magnetoresistance is observed and the Hall resistivity follows linear field dependence. Upon cooling down the system below $T_s$, large magnetoresistance develops and the Hall resistivity deviates from the linear field dependence. Furthermore, we found that the transition at $T_s$ is extremely robust against the external magnetic field. We argue that the magnetic state in CeFeAsO is unlikely a conventional type of spin-density-wave (SDW).

## 1. Introduction

Following the discovery of superconductivity in $La(O_{1-x}F_x)FeAs$ [1], variant families of the iron-based superconductors have been successively found in the last two years. Resembling the high $T_c$ cuprates, all these iron pnictides crystallize in a layered crystal structure. Superconductivity occurs upon suppressing the magnetic transition via either doping [2] or applying pressure [3], which provided another important example for studying the interplay of superconductivity and magnetism. On the other hand, the iron pnictides also exhibit many distinct features from the high $T_c$ cuprates. For example, superconductivity in iron pnictides is nearly isotropic even though they are layered crystals [4,5], and their parent compounds are usually bad metal instead of a Mott insulator [6]. A comparison of these two types of superconductors might provide important insights on the puzzles of high $T_c$ superconductivity.

In order to study the interplay of superconductivity, magnetism and structural distortion and, therefore, the mechanism of superconductivity in iron pnictides, it is crucial to elucidate the nature of the magnetic state/structural phase transition in the parent compounds. Variant approaches, either based on Fermi surface nesting [7,8] or started from the proximity to a Mott-insulator [9], have been proposed to describe the magnetic states, but no consensus has been reached. A fundamental problem concerns on the itinerancy of the 3d electrons, which degree can be measured against the robustness of the magnetic transition to external magnetic field. Measurements in a pulsed magnetic field are, therefore, highly desired. In iron pnictides, intensive efforts have been devoted to the superconducting properties using pulsed magnetic field [4], but little attention has been drawn into the magnetic state.

In this paper, we report the measurements of the electrical resistance $R_{xx}(\mu_0H)$ and the Hall resistivity $R_{xy}(\mu_0H)$ of CeFeAsO up to 45T. We found that both $R_{xx}(\mu_0H)$ and $R_{xy}(\mu_0H)$ demonstrate distinct behaviour in the magnetic and paramagnetic state, respectively. The observations of extreme robustness of the magnetic transition/structural transition to the applied magnetic field might indicate that the magnetic state is unlikely a conventional type of spin-density-wave state.

## 2. Experimental methods

Polycrystalline samples of CeFeAsO are synthesized by solid state reaction method using CeAs, $CeO_2$, $Fe_2As$ as starting materials [10]. The raw materials were thoroughly mixed and pressed into pellets, which were then wrapped with Ta foil and sealed in an evacuate quartz tube. The materials were then annealed at 1150 °C for 50 h. The resulting samples were characterized by powder x-ray diffraction (XRD) at room temperature, which identifies a single phase with a tetragonal ZrCuSiAs-type structure (space group *P4=nmm*).

The longitudinal resistivity and the transverse Hall resistivity were simultaneously measured with a typical 5-probe method using the facilities of pulsed magnetic fields in the National High Magnetic Field Laboratory at Los Alamos. In order to eliminate the effects of contact asymmetries, forward- and reverse-field shots were made at the same temperatures for Hall resistance measurements. The data traces were recorded on a digitizer using a custom designed high-resolution, low-noise synchronous lock-in technique. The temperature dependence of the resistivity at zero magnetic field was measured with a Lakeshore resistance bridge. To minimize the self-heating effect in a pulsed magnetic field, samples with typical size of about 2mm×0.5mm×0.02mm were cut off from the large batch for resistivity measurement.

## 3. Results and discussion

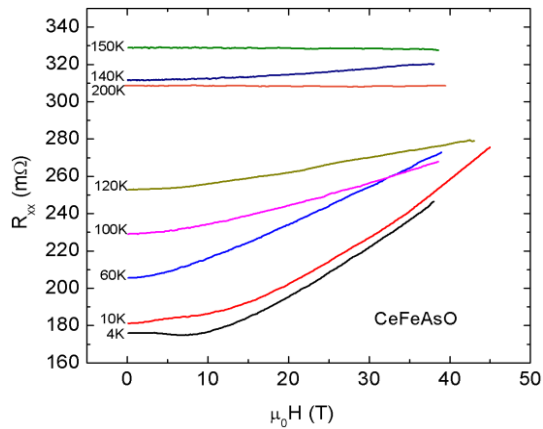

**Figure 1:** The field dependence of the electrical resistance $R_{xx}(\mu_0H)$ for CeFeAsO at variant temperatures of T=4K,10K,60K,100K,120K,140K,150K, and 200K, respectively.

Figure 1 shows the electrical resistance $R_{xx}(\mu_0H)$ of CeFeAsO at variant temperatures. It is noted that CeFeAsO undergoes a structural phase transition around $T_s \approx 150K$, followed by an antiferromagnetic transition around $T_N \approx 140K$ as derived from the measurements of neutron scattering [11]. The difference between $T_s$ and $T_N$ might become smaller with improving sample quality [12]. From figure 1, one can see that the magnetoresistance is negligible at temperatures above the structural phase transition (T>150K), but becomes significant once cooling down below $T_s$. At low temperature (e.g., T=4K), a large magnetoresistance is observed.

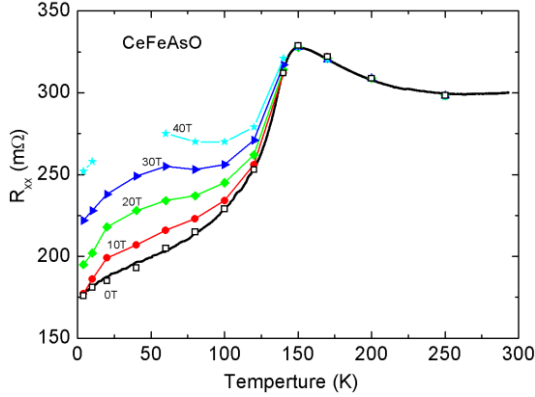

**Figure 2.** Temperature dependence of the electrical resistivity Rxx(T) at variant magnetic fields for CeFeAsO. The thick line shows the zero-field resistivity measured while cooling down the sample, and the symbols represent the data measured at variant.

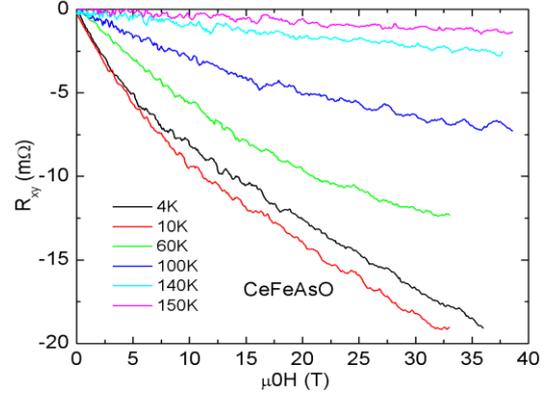

**Figure 3.** The Hall resistance $R_{xy}(\mu_0 H)$ of the CeFeAsO at variant temperatures.

To show the evolution of the magnetic/structural transitions with magnetic field, the electrical resistivity is replotted as a function of temperature in figure 2. Note that the thick line is the resistive curve measured with the Lakeshore resistance bridge while cooling down the sample at zero field. Obviously, the data collected at variant shots (empty squares) are consistent with those measured at zero field. From figure 2, one can see that the electrical resistivity, $R_{xx}$, is largely enhanced at low temperatures upon applying a magnetic field, but the magnetoresistance seems to always vanish around 150K at which a phase transition from tetragonal to orthorhombic structure has been identified [11,12]. The disappearance of magnetoresistance at the structural phase transition suggest that spin fluctuations or short range magnetic order might already develop at the structural phase transition, prior to the magnetic transition. This indicates the importance of magnetic interactions on the subsequent occurrence of the structural and magnetic phase transitions in CeFeAsO. Remarkably, the structural phase transition is independent of the magnetic field up to 40T. Since the structural transition is likely driven by magnetic interaction, the extreme robustness of the structural transition to magnetic field demonstrates that the magnetic state in CeFeAsO cannot be a conventional type of spin-density-wave state. Otherwise, the magnetic/structural transitions would be suppressed by applying a sufficiently large magnetic field. A magnetic field induced shoulder-like transition is observed with further lowering temperature which nature remains unclear.

Figure 3 shows the field dependence of the Hall resistance at variant temperatures. It is found that the behaviour of the Hall resistance $R_{xy}(\mu_0 H)$ also undergoes a dramatic change at the structural phase transition ($T_s$=150K). At temperatures above $T_s$, the Hall resistivity follows linear field dependence, but deviates from such linear behaviour as the system is cooled below $T_s$. Band structure calculations show two types of charge carriers (holes and electrons) in iron pnictides[13]. According to the two-band model[14], the linear field dependence of $R_{xy}(\mu_0 H)$ hints a good balance of electrons and holes above $T_s$. The deviation of the Hall resistivity from the linear field dependence below $T_c$ indicates that the charge carriers or Fermi surface is significantly modified in the magnetic state. As seen in figure. 3, the initial slopes of the Hall resistivity decrease faster at temperatures below $T_s$, suggesting a loss of electronic carriers below the structural phase transition.

The above experimental findings, including the magnetoresistance and the Hall resistance, strongly indicate that the electronic properties of CeFeAsO undergo substantial changes at the structural/magnetic transitions, which is likely driven by magnetic interactions. To elucidate the nature of the magnetic transition and its interplay with the structural transition, it is desired to study the single crystals and to extend the measurements to higher magnetic fields.

## 4. Conclusion

To summarize, we have found that both the magnetoresistance and the Hall resistivity of CeFeAsO show clear changes at the structural phase transition. Independent of the applied magnetic field, the magnetoresistance always vanishes at the structural transition, indicating a close relation between the structural distortion and the magnetic ordering, which might originate from the magnetic interactions. The steeper slopes of the Hall resistivity $R_{xy}(\mu_0 H)$ and its deviation from the linear field dependence at temperatures below $T_s$ indicate that the magnetic/structural transitions might lead to a change of the electronic structure.

**Acknowledgements**
We acknowledge very helpful discussions with J. P. Hu, T. Xiao, Z. Y. Weng and M. B. Salamon. This work was supported by NSFC, the National Basic Research Program of China (973 program), the PCSIRT of the Ministry of Education of China, Zhejiang Provincial Natural Science Foundation of China and the Fundamental Research Funds for the Central Universities. Work at NHMFL-LANL is performed under the auspices of the National Science Foundation, Department of Energy and State of Florida.